# Polarization multiplexed dissipative Kerr solitons in on-chip micro-resonator

YONG GENG[1], YANLAN XIAO[1], XINJIE HAN[1], KUN QIU[1], JING XU[2] AND HENG ZHOU[1, *]

[1]Key Lab of Optical Fiber Sensing and Communication Networks, School of Information and Communication Engineering, University of Electronic Science and Technology of China, Chengdu 611731, China
[2]Wuhan National Laboratory for Optoelectronics, Huazhong University of Science and Technology, Wuhan 430074, China
*Corresponding author: zhouheng@uestc.edu.cn

**We demonstrate polarization multiplexed dissipative Kerr solitons in an on-chip silicon nitride micro-resonator. In our experiment, TE- and TM-polarized soliton can be individually generated and controlled, thanks to their weak mutual interaction as the result of sufficiently different repetition rates and orthogonal polarization states. Furthermore, we find that TE- and TM-polarized solitons usually exhibit uncorrelated time jitters, therefore the frequency and phase coherence between the polarization multiplexed soliton microcombs change dramatically as a function of pump laser parameters, by optimizing which we achieve narrow dual-microcomb beat note linewidth as small as 4.4 kHz. Potential applications of on-chip polarization multiplexed soliton microcombs include Kerr comb spectral expansion, dual-comb metrology, and measurement of quantum entanglements.**

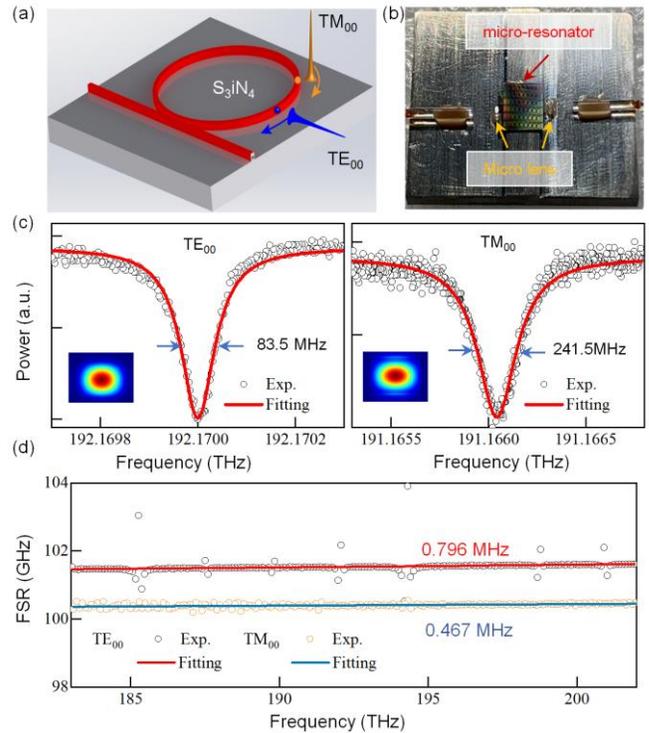

**Fig. 1** Polarization multiplexed DKSs formed in an $Si_3N_4$ micro-resonator. (a) The illustration of polarization multiplexed DKSs. (b) The picture of packaged $Si_3N_4$ chip used in the experiment. (c) Measured linewidth for two $TM_{00}$ and $TE_{00}$ cavity resonances; The inset shows the simulated mode profile. (d) Measured FSR for mode $TM_{00}$ and $TE_{00}$.

Dissipative Kerr soliton (DKS) has received significant research interest thanks to its performance merits for various applications including wavelength division multiplexed (WDM) optical communications, precision laser spectroscopy, and optical frequency synthesize [1-3]. Generally, DKS can be generated in externally driven optical micro-resonators when the double-balance between Kerr nonlinear phase shift and cavity dispersion, as well as between Kerr parametric gain and cavity decay are achieved, giving rise to ultra-short mode-locked pulse trains, and broadband low noise Kerr frequency combs (soliton microcombs) [4, 5]. Specifically, in some application scenarios such as Vernier spectrometer, dual-comb spectroscopy and ranging, broadband signal channelization, it requires more than one DKS microcombs with properly differed repetition rates [6-9], motivating the generation of two or more soliton microcombs within one micro-resonator. To date, multiple DKS microcombs generation has already been demonstrated in micro-resonators via direction multiplexing [10], spatial (mode) multiplexing [11], and spectral multiplexing [12]. However, as an elementary multiplexing technology in fiber optics [13], polarization multiplexing of soliton microcombs within on-chip micro-cavity has not been experimentally demonstrated [14], mainly due to the sizable cavity thermal nonlinearity and close interactions of solitons that hinder the reliable and simultaneous access of TE- and TM-solitons.

In this letter, we demonstrate that a dual-pump driven $Si_3N_4$ micro-resonator can support stable generation and operation of two polarization multiplexed DKSs, assisted by the method of auxiliary laser heating (ALH) method [15]. Due to that TE- and TM-polarized solitons are generated in orthogonal polarization modes of the micro-resonator with different free spectral range (FSR), we show that the orthogonally polarized DKSs can be individually generated and controlled, and they usually exhibit uncorrelated repetition rate fluctuations, therefore the phase coherence between the polarization multiplexed soliton microcombs is highly dependent on the pump laser parameters. By using a coherent pumping scheme with optimized pump laser configuration, we achieve the narrowest beat note linewidth of 4.4 kHz between two orthogonally polarized soliton microcombs.

Figure 1(a) illustrates the concept for the generation of polarization multiplexed DKSs in a single $Si_3N_4$ micro-resonator. Figure 1(b) shows the picture of packaged $Si_3N_4$ chip used in our experiment. The cross-section of this micro-resonator is 1650×800 $nm^2$ and the FSR is around 100 GHz. The $Si_3N_4$ chip is packaged with UV-glued I/O lenses and fibers that can effectively suppress the random fluctuation of on-chip pump power [16]. To warrant DKS formation, both the TE and TM mode of the micro-resonator should have high enough Q-factor and anomalous dispersion. Hence, we use the Mach–Zehnder interferometer (MZI)-based optical sampling technique to measure the FSR and chromatic dispersion [17]. Figure 1(c) exemplifies the measured full width at half maximum (FWHM) of two $TE_{00}$ mode and $TM_{00}$ mode resonances, showing a corresponding to a loaded Q-factor of $2.3×10^6$ and $7.9×10^5$, respectively. Moreover, in order to identify the resonator dispersion, we record the resonance frequencies of the cold micro-resonator between 1480nm and 1640nm, calibrated by the MZI sampling technique. The extracted FSRs for both $TE_{00}$ and $TM_{00}$ mode are plotted in Figure 1(d), it is seen that their FSRs have different mean values of 101.53 GHz ($TE_{00}$) and 100.38 GHz ($TM_{00}$), but both featuring anomalous dispersion, i.e., 0.796 MHz for the $TE_{00}$ and 0.467 MHz for the $TM_{00}$ respectively [18].

The experiment setup used to generate polarization multiplexed DKSs is illustrated in Figure 2(a). Two individual external cavity lasers, which are centered at 1602.8 nm and 1547.9nm, are used as TE and TM mode pump lasers respectively. The polarization states of these two pump lasers are adjusted by using two fiber polarization controllers (FPC) to align with the TM and TE mode of the micro-resonator respectively. Then the TE- and TM-polarized pump lasers are boosted to 32.5 dBm and 32.0 dBm using two high-power Er-doped fiber amplifiers (EDFA) and combined by using a 25:75 optical coupler and launched into the resonator chip, setting the on-chip pump power for TE- and TM-polarized mode to about 23.5 dBm and 27.8 dBm. To suppress cavity thermal nonlinearity, an auxiliary light from another external cavity laser is amplified to 32 dBm and launched into one resonance of TM mode near 1535.0 nm for the counterpropagating direction (see Fig. 2a). As the auxiliary laser enters the cavity mode from the blue detuning range, the micro-resonator is heated, and all the resonances are thermally redshifted, letting the TE- and TM-polarized pump lasers passively enter the resonances from the red detuning range. By properly setting the frequency detuning between the two pump lasers, the heat flow can be well balanced out and the pump laser can scan across the entire resonance without notable thermal dragging.

Figure 2(b) shows typical optical spectra of polarization multiplexed soliton microcombs measured at the output of the silicon nitride chip. In our experiment, by individually adjusting the pump detuning parameters for the TE and TM polarized pump laser, different soliton composites can be generated within the micro-resonator, as presented in Figure 2(b). Also, it is seen that the orthogonally polarized soliton microcombs illustrate quite different spectral envelopes from each other, due to the non-uniform Q-factor, chromatic dispersion and bus-to-resonator coupling coefficient between the corresponding TE and TM modes. In particular, the TM-polarized DKS shows a 20-dB spectral bandwidth of ~98 nm (mainly relying on the much higher Q-factor), while the 20-dB bandwidth of TE-polarized DKS spectrum is just ~58

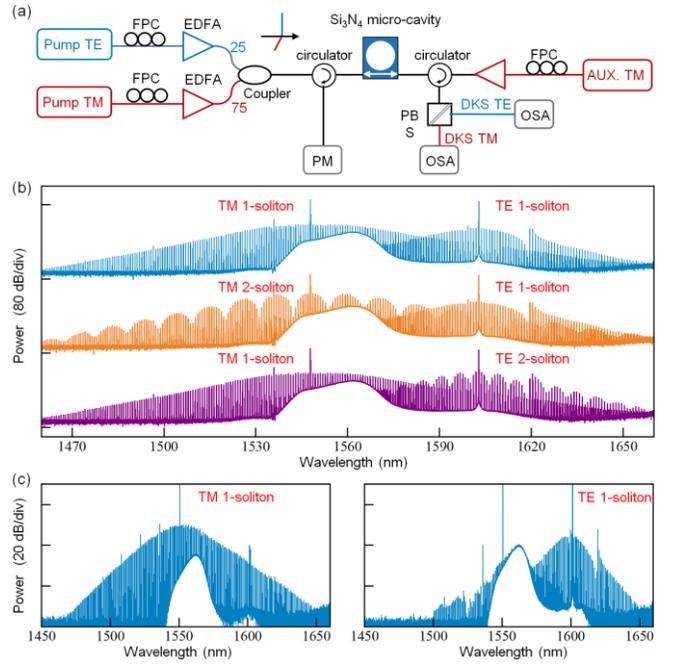

**Fig. 2** (a) Experimental setup for polarization multiplexed soliton microcombs generation using ALH method. (b) Obtained dual orthogonally polarized soliton microcomb spectra. (c) The spectrum of separated TM- and TE-polarized single soliton using a PBS. FPC, fiber polarization controller; EDFA, erbium-doped fiber amplifier; PM, power meter; OSA, optical spectrum analyzers; PBS, polarization beam splitter.

nm. Furthermore, at the output of the silicon nitride micro-resonator, we can separate the polarization multiplexed soliton microcombs by using a polarization beam splitter (PBS), as shown in Figure 2(c), confirming that the microcombs are indeed of orthogonal polarization. Such separability makes the polarization multiplexed microcombs a truly dual-comb source that is applicable for functionalities such as phase-sensitive dual-comb spectroscopy and high-speed signal channelization.

Besides, Figure 3 shows the process that we generate the multiplexed TM and TE- soliton microcombs. In our experiment, the TM-polarized soliton microcomb centered around 1550 nm is first generated (see panel i in the Figure 3), then we tune another TE-polarized pump laser into another cavity resonance belonging to TE mode family (centered at ~1600 nm). When we scan the TE-polarized pump laser from the blue detuning to red detuning range, it goes through the stages of low-noise primary comb line state (panel ii), high noise chaos state (panel iii and iv), multiple soliton state (panel v) and finally reaches the single soliton state (point vi), as shown in Figure 3. Intriguingly we see that during the evolution of TE-polarized Kerr comb, the already existed TM-polarized soliton microcomb remains nearly unimpacted. Such weak interaction between the polarization multiplexed microcombs can be attributed to the following reasons. First, thanks to the auxiliary heating method, the thermal shift of cavity resonance caused by the TE-polarized pump is negligible as the TE-pump power is 6dB smaller than the TM-pump laser. Second, since the TE- and TM- pump lasers are orthogonally polarized, the nonlinear interaction between them via four-wave mixing (FWM) is prohibited, so they can operate independently. Third, as shown in Figure 1d, the measured TE mode FSR is $f_{TE}$ = 101.53 GHz while the TM mode FSR is $f_{TM}$ =100.38 GHz, which imply that the TE and TM solitons have different group velocities. After being generated within the micro-cavity, they walk off with each other quickly, and therefore they do not have enough time to impose strong perturbation (e.g., via Kerr cross phase modulation) to each other. This is distinct from the bind solitons in the scenarios of spatial multiplexing reported in in prior literature [11]. To further validate this phenomenon, in a new experiment, we generate polarization multiplex solitons with different cavity mode distances between the TE- and TM-polarized pump laser as shown in Figure 4. The experimental results confirm that, even adjacent TE and TM modes are pumped (i.e., the case labeled as 0×FSR in Figure. 4), orthogonally polarized solitons can still be generated, whose spectra are fully overlapped but still exhibit unobvious mutual interaction.

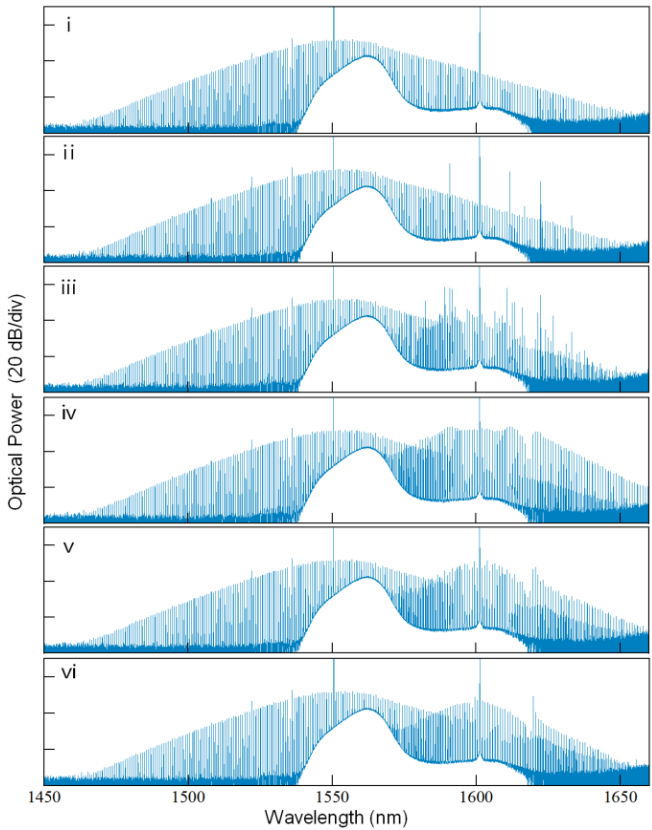

**Fig. 3** The microcomb spectra generated with polarization multiplexed dual-pump configuration. The TE-polarized microcomb can reach different state from turning pattern to single soliton state, while the TM-polarized microcomb maintains single DKS state.

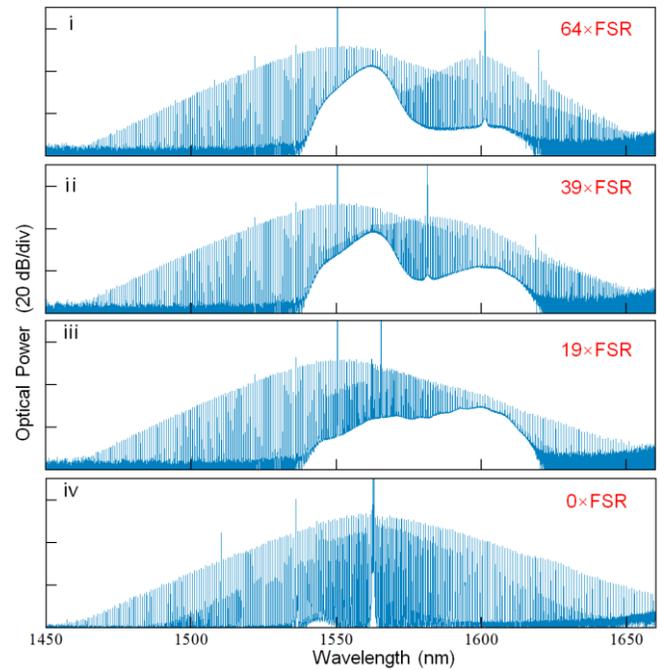

**Fig. 4** The microcomb spectra with different mode number spacing between TE- and TM-polarized pump.

The optical spectra shown in Figure 3 and Figure 4 imply abundant potential applications of polarization multiplexed Kerr microcombs. For example, the spectra in Figure 4 panel i and panel ii indicate that the TM and TE soliton microcomb can respectively cover the C- and L-band in the scenarios of WDM optical communications, bringing about laser tones with higher power levels in comparison with either a single TE or TM microcomb. Besides, the combination of chaotic TE comb and soliton TM comb shown in panel iv Figure 3 provides an ideal dual-comb source for chaotic laser ranging proposed recently [19], the chaotic TE comb lines can be used to probe the target while the mode-locked TM comb lines can be used to retrieval the ranging and velocimetry information. Moreover, the overlapped TE and TM soliton microcombs shown in panel v Figure 4 is applicable for wideband signal channelization [20], considering their ~1 GHz mode spacing discrepancy. Importantly, for signal channelization function, the frequency and phase coherences between the polarization multiplexed soliton microcombs should be as high as possible, so that to avoid possible signal distortion. So next we will study how to enhance the coherence between the TE and TM soliton microcombs.

In this experiment, stead of using two independent lasers, we use an electro-optic modulation to produce both the TE and TM pump laser tones. Particularly, an external cavity semiconductor laser operating at 193.866 THz is split via a 50:50 optical coupler, one part is used as the TM pump to generated TM-polarized soliton, and the other part is sent into an optical amplitude modulator to produce a proper frequency shift and then serve as the TE pump. Here, the amplitude modulator is driven by a microwave synthesizer operating at 17.506 GHz, and the third-order modulation sideband is filtered out and used as the TE pump to generate TE-polarized DKS (about 52.52 GHz offset from the original tone, so as to meet the mode spacing between TE and TM mode resonance). The generated optical spectrum is shown in the Figure 5(a), and the inset shows the dual-comb beat notes between the orthogonally polarized soliton microcombs, revealing a mode spacing discrepancy of about 1.1 GHz.

Furthermore, Figure 5b summaries the 25th inter-comb beat note linewidth, which changes from a few hundred kilohertz to the narrowest value of 4.4. kHz under different TE pump detuning values. It is now well known that the dissipative Kerr soliton repetition rate is a function of pump-cavity detuning, intermediated by the Raman-induced soliton-self-frequency shift (SFFS) and dispersive-wave induced spectral recoil [21-23]. So, the phenomenon illustrated in Figure 5(b) can be attributed to the uncorrelated repetition rate jitters between the TE and TM polarized DKS, since the TE and TM mode DKS shows quite different spectrum envelope, and hence prone to experience disparate Raman and dispersive-wave induced spectrum drift. Therefore, although the TE and TM pump is derived from a same external cavity laser and thus undergo identical frequency fluctuation, their repetition rate fluctuations can be different. At those optimized detuning values (point iv, v and vi in Figure 5(b)), the TE and TM solitons share the most similar slope on the repetition rate versus pump detuning curve, thus giving rise to narrowest inter-comb beat note (i.e., about 4.4 kHz at point v) and strongest coherence between the TE and TM soliton microcombs. It is worth noting that in the above experiment, we did not observe prominent injection locking behavior between the TE and TM solitons [10]. As can be seen from Figure 5a, when the TE and TM microcomb lines encounter each other (prerequisite for injection locking) around 1520nm and 1625 nm, where the comb line powers have already become quite low (especially for the TE microcomb) and might be the reason hindering injection locking. Nevertheless, if and how we can achieve all-optical locking between polarization multiplexed soliton microcombs needs further investigation.

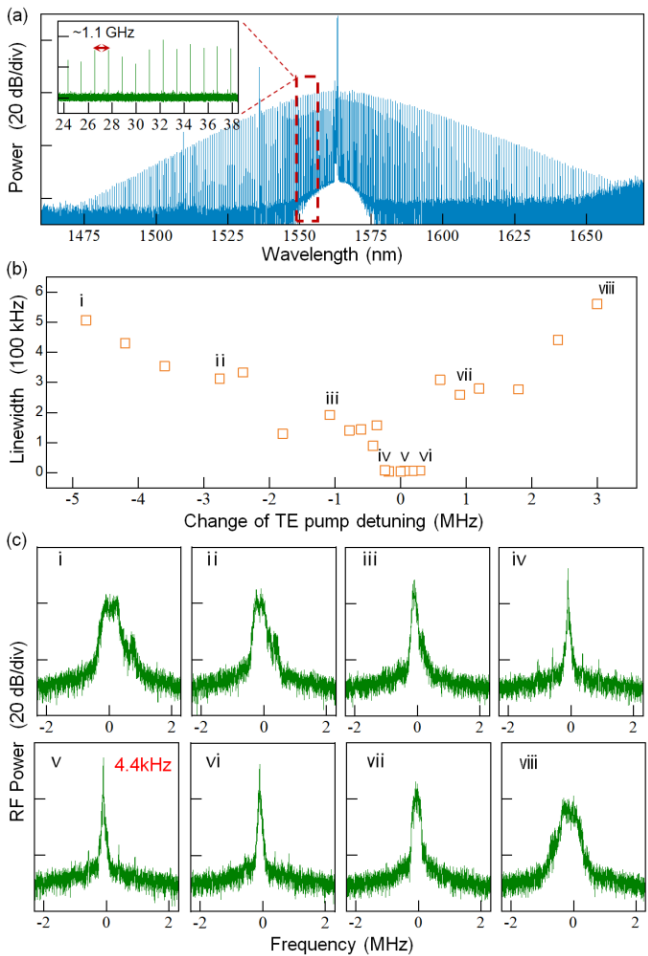

Fig. 5. (a) Optical spectrum of orthogonally polarized DKSs. The inset is the dual-comb beat note. (b) Measured linewidth of dual-comb beat note versus TE mode pump laser-cavity detuning. (c) Electrical spectrum of dual-comb beat note with different TE pump laser-cavity detuning.

In conclusion, we demonstrated polarization multiplexed solitons generated in an on-chip silicon nitride micro-resonator. Because such orthogonally polarized DKSs characterizing a sufficiently different repetition rate and orthogonally polarized state, the soliton composite revealed weak mutual interaction, enabling individually generation and manipulation. Furthermore, we experimentally illustrated that the frequency and phase coherence between the TE- and TM-polarized microcomb lines can be strengthened by optimizing the pump laser parameters. Our study offers a low-complexity, high robust solution to generate separable dual-microcomb, which hold great potentials in applications including Kerr comb spectral expansion, dual-comb spectroscopy and ranging.

**Funding.** This work is supported by the National Key Research and Development Program of China (No. 2019YFB2203103), the NFSC grant (No. 62001086 and 61705033).